# Interplanetary Radio Emission: A Summary of Recent Results

Nat Gopalswamy

Solar Physics Laboratory, NASA Goddard Space Flight Center, Greenbelt, MD 20771, USA

**Abstract**

This paper summarizes some recent results in the low-frequency radio physics of the Sun. The spatial domain covers the space from the outer corona to the orbit of Earth. The results obtained make use of radio dynamic spectra and white-light coronagraph images and involve radio bursts associated with solar eruptions and those occurring outside solar eruptions. In particular, the connection between type II radio bursts and the sustained gamma-ray emission from the Sun is highlighted. The directivity of interplanetary type IV bursts found recently is discussed to understand the physical reason behind it. A new event showing the diffuse interplanetary radio emission (DIRE) is introduced and its properties are compared with those of regular type II bursts. The DIRE is from the flanks of a CME-driven shock propagating through nearby streamer. Finally, a new noise storm observed by two spacecraft is briefly discussed to highlight its evolution over two solar rotations including the disruption and recovery by solar eruptions.

## 1. Introduction

Both thermal and nonthermal radio emission from the Sun are due to electrons. Thermal emission provides information on the quiet Sun and prominences. Nonthermal emission reflects energy release processes on the Sun that energize electrons. Energetic electrons emit radio emission through various processes such as gyro synchrotron and plasma emission depending on the magnetic and plasma environment into which these particles are released. The processes such as magnetic reconnection and fast-mode MHD shock that accelerate electrons can also accelerate ions that escape into interplanetary space to be observed as solar energetic particles (SEPs) or precipitate into the solar atmosphere and produce gamma-rays via different processes. Nonthermal electrons also produce X-rays and gamma-rays via bremsstrahlung. Thus, the Sun is an important laboratory to study various nuclear, atomic, and magneto-plasma processes (see reviews by Benz 2017; Vilmer 2012; Gopalswamy 2011).

Radio emission at frequencies below the ionospheric cutoff (~15 MHz) need to be observed from space. The low-frequency radio emission is primarily due to plasma emission manifesting as radio bursts of type II, type III, and type IV. These bursts are related to large solar eruptions that involve intense flares and coronal mass ejections (CMEs) and hence are eruption bursts. Type III bursts are the most common ones as they occur also during less energetic eruptions, such as jets. There is also weaker emission generally referred to as radio noise storms but manifest as a series of short-duration type III bursts in quick succession emitted over many days. The low-frequency emission has also important implications for space weather because they identify CMEs that result in large SEP events and intense geomagnetic storms. Although the low-frequency radio emission has been observed since the 1970s at frequencies below 2 MHz, the crucial gap between 2 MHz and the ionospheric cutoff was filled by the Radio and Plasma Wave Experiment (WAVES) on board Wind (Bougeret et al. 1995) and the Solar Terrestrial Relations Observatory (STEREO, Bougeret et al. 2008). These instruments have been providing information on low-frequency radio emission from the Sun since 1994 and 2006, respectively. Alone and together, these observations have resulted in a number of discoveries, thanks to the simultaneous availability of coronagraphs with overlapping spatial coverage on STEREO and the Solar and Heliospheric Observatory (SOHO) as summarized in Gopalswamy (2011).

This paper is concerned with selected recent results on the low-frequency radio emission that include both eruption bursts and type III storms. These results highlight the importance of combining radio observations with those from other wavelengths from white-light to gamma-rays.

## 2. Interplanetary type II bursts

Interplanetary (IP) type II bursts indicate that shocks driven by CMEs accelerate particles throughout the inner heliosphere and the energetic electrons from the shocks result in the radio burst via the plasma emission mechanism. Wind/WAVES has enabled tracking these shocks from the corona to 1 AU (Reiner et al. 1999; Liu et al. 2013; Cremades et al. 2015; Gopalswamy et al. 2005; 2018a). Only tens of type II bursts were studied in the pre-Wind era, whereas now we have several hundred of them observed and cataloged (Gopalswamy et al. 2019a). Here we highlight two results on the possible locations on the shock surface that radiate the bursts and the type II burst connection to sustained gamma-ray emission (SGRE) from the Sun.

### 2.1 Where does a type II burst originate on the shock surface?

Both shock nose and flanks have been proposed as sites of electron acceleration in CME-driven shocks (Gopalswamy 2011; Magdalenić et al. 2014; Gopalswamy et al. 2018a,b; Krupar et al. 2019). The location where the emission comes from depends whether the shock is strong enough to accelerate electrons that produce type II radio bursts via the plasma emission mechanism at the fundamental and harmonic of the local plasma frequency. The CME-driven shock is curved and hence cuts through many density layers and different magnetic field geometries at a given time. These layers have different plasma frequencies, so the radio emission can occur at different frequencies. The flank is at a lower height hence at a higher density and plasma frequency. The shock speed is at its maximum in the nose region, decreasing monotonically toward the flanks. The Alfven speed in the corona has a minimum value at ~1.5 Rs, attains a peak around 3 Rs and then decreases with distance. Thus, a combination of shock speed and the Alfven speed profile determines the shock strength. Since metric type II bursts occur at a heliocentric distance <2 Rs, the flanks usually are preferred for the type II burst. Also, the fact that CMEs have similar speeds in the radial and lateral directions early on supports this. As a CME travels into the IP medium, the speed becomes more radial, so the shock flanks weaken, while the nose strengthens making it possible for type II emission to occur there.

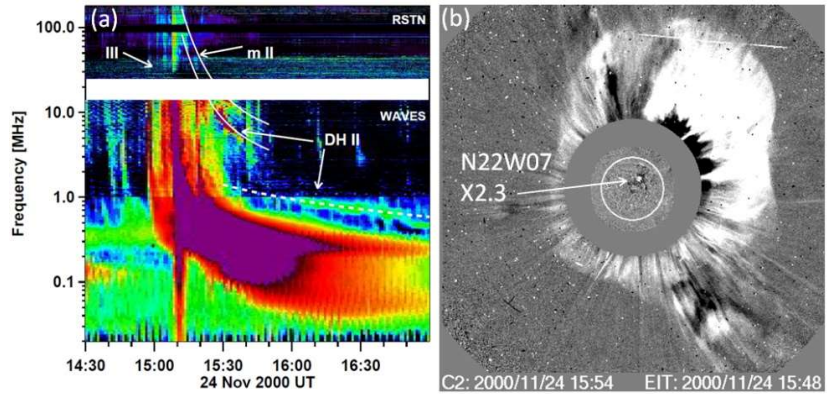

Figure 1. A composite dynamic spectrum from Radio Solar Telescope Network (RSTN) and Wind/WAVES showing type II burst components (a) during the 2000 November 24 CME (b). The high-frequency component extends from metric (m) to decameter-hectometric (DH) wavelengths (traced by the solid white curves: upper - fundamental and lower - harmonic). The low-frequency component starts at DH and extends to kilometric (km) wavelengths (traced by the dashed line). Note that there is no frequency relationship between the m-DH and DH-km components of the type II burst. The CME was a full halo with the flux rope part moving predominantly in the northwest direction. The CME originated from close to the disk center (N22W07) in association with an X2.3 flare.

Figure 1 shows the type II burst associated with the 2000 November 24 fast (~1500 km/s) halo CME from close to the disk center (https://cdaw.gsfc.nasa.gov/CME_list/halo/halo.html). The CME was at a sky-plane height of ~6 Rs when the m-DH component ended, and the DH-km component started around 15:50 UT. The m-DH component has fundamental-harmonic structure (white lines in Fig 1a), while the DH-km component has does not have that structure. At ~15:50 UT, the DH-km burst is at 1 MHz, whereas the m-DH fundamental is at 4 MHz. Thus, the frequency of the m-DH component is a factor of 4 higher (factor of 8 if the DH-km is harmonic), indicating that the plasm density is smaller by a factor of 2-3 at the nose than that in the flanks.

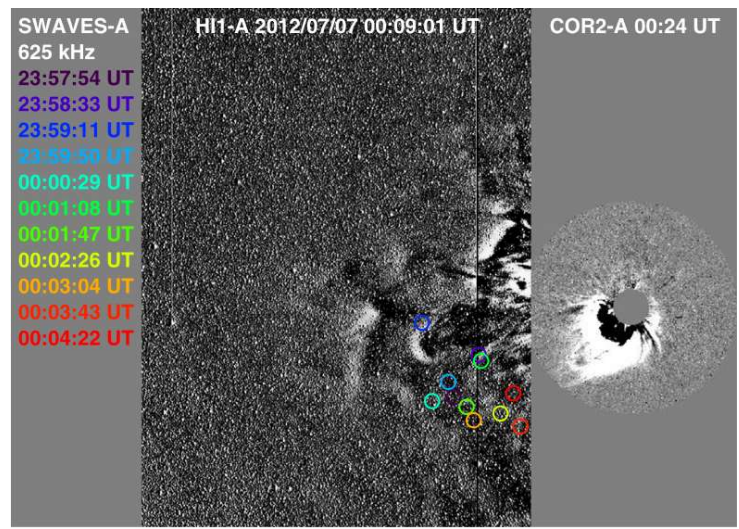

Figure 2. DH type II source locations at 0.625 MHz (colored circles corresponding to various times) over-plotted on the CME image from STEREO-A Heliospheric Imager 1 (HI1) and COR2 running difference images. The radio sources are located above the nose of the CME heading in the southeast direction (Mäkelä et al. 2018).

During a similar event on 2012 July 6, the m-DH component started in the metric domain (~80 MHz), ending around 8 MHz; the DH-km component started around 1 MHz and ended around 0.4 MHz (Mäkelä et al. 2018). The m-DH and DH-km components overlapped in time for more than an hour. The CME was observed by SOHO and STEREO coronagraphs and heliospheric imagers. The direction-finding analysis at 625 kHz from STEREO/WAVES places the DH-km radio sources close to the nose region of the associated CME as seen in Fig. 2. However, the radio source locations were shifted away from the Sun and the leading edge of the associated CME, probably due to the scattering of the radio waves. A similar conclusion was reached when the source locations obtained from Wind/WAVES are plotted on SOHO/LASCO images (not shown).

## 2.2 Type II Radio Bursts and SGREs

Gamma-ray emission from the Sun lasting beyond the impulsive phase of the associated flares is primarily due to the decay of neutral pions (Forrest et al. 1985). The production of neutral pions requires >300 MeV protons precipitating into the chromosphere from the acceleration site in the corona. The key finding of Forrest et al. is the time-extended nature of the gamma-ray emission and hence called a long-duration gamma-ray flare (LDGRF). Ideas focusing on the time-extended nature of these gamma-rays looked for ways to extend the life of the >300 MeV protons accelerated in the impulsive phase, e.g., trapping them in flare loops (e.g., Ryan and Lee 1991). The extended-phase emission can also be explained by protons diffusing toward the photosphere from coronal shocks, while the impulsive phase emission is due to protons from the flare (Murphy et al. 1987). The Large Area Telescope (LAT) on board the Fermi satellite has detected many such events, some of them lasting for almost a day. These are now termed sustained gamma-ray emission (SGRE) to note that the emission continues long after the end of the associated flare (Plotnikov et al. 2017; Gopalswamy et al. 2018c) or late-phase gamma-ray emission (LPGRE, Share et al. 2018). Share et al. (2018) noted that most of the SGREs are associated with fast CMEs (speed >800 km/s) and IP type II bursts. Plotnikov et al. (2017) had noted the association with metric type II bursts in the three cases they studied.

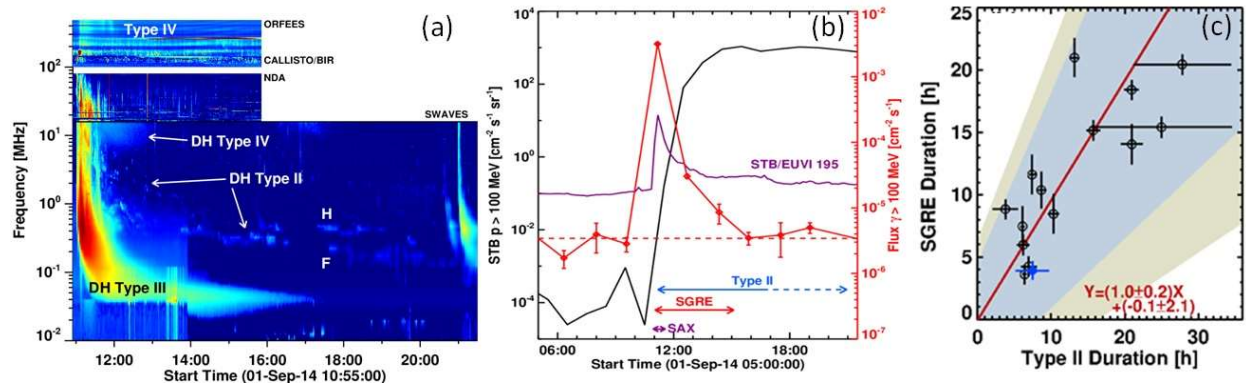

Figure 3. (a) STEREO/WAVES dynamic spectrum showing the type IV, Type II, and type III bursts associated with the 2014 September 1 SGRE event. The metric radio data are from Nancay Decametric Array (NDA) and CALLISTO. (b) >100 MeV gamma-ray flux (red), >100 MeV proton intensity from STEREO particle detectors (black), and EUV flare intensity from STEREO-B/EUVI (purple); the type II burst and soft X-ray flare (SAX) durations are marked. (c) Scatterplot between gamma-ray duration and type II burst duration for 19 SGRE events with the blue data point showing the 2014 September 1 event (adapted from Gopalswamy et al. 2019b; 2020a).

An important breakthrough in the association between SGRE and shocks came about when it was found that the SGRE duration and IP type II duration are linearly related (Gopalswamy et al. 2018c; 2019b; 2020a). The relation between SGRE events and CME-driven shocks inferred from IP type II bursts offers a convincing evidence for the shock origin of >300 MeV protons required for producing an SGRE event. SGRE events have been observed by Fermi (in Earth orbit) from some backside solar eruptions (Pesce-Rollins et al. 2015). For example, the 2014 September 1 event was ~36º behind the limb yet produced a 4-hr duration SGRE detected by Fermi/LAT. Fortunately, this event was a disk event for STEREO-B, which provided details on the spatial structure of the eruption (post-eruption arcade and twin dimming regions). STEREO/WAVES observed an IP type II burst that lasted >7 hrs (see Fig. 3a) associated with an ultrafast CME (>2000 km/s) observed by STEREO and SOHO. The >100 MeV gamma-rays extended well beyond the soft X-ray peak inferred from STEREO-B/EUVI (Fig. 3b).

The time profile of >100 MeV proton intensity derived from STEREO-B particle data (Fig. 3b) increased by >4 orders of magnitude during the SGRE indicating the presence of >300 MeV protons required for the SGRE. The durations of the STEREO type II burst (about 7.5 hours) and SGRE (~4 hrs) are in good agreement with the linear relation (Fig. 3c) found between the two durations (Gopalswamy et al. 2019b). In the shock scenario, the gamma-ray source is spatially extended because the angular extent of the shock is much larger than that of the flare structure. The shock occupies a large volume indicating a spatially extended accelerator, suggesting that the SGRE source at the Sun is likely to be spatially extended (see Cliver 1993 for a similar scenario but for gamma-ray line emission).

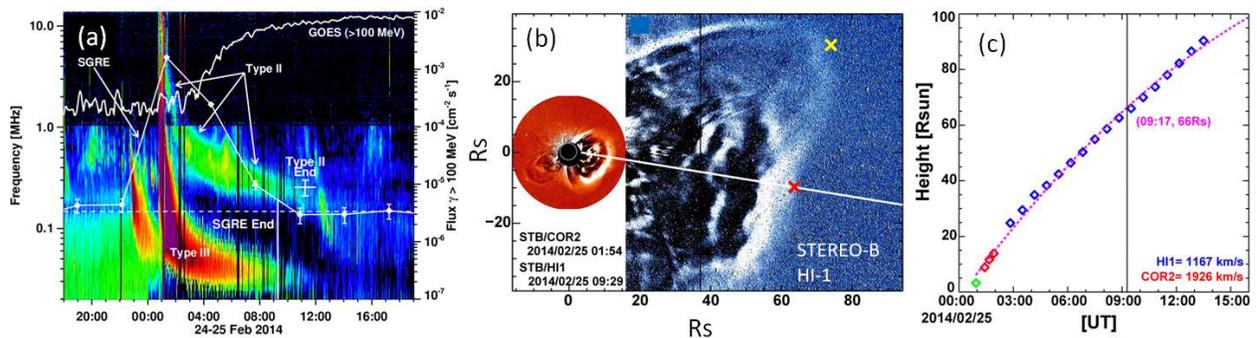

Figure 4. (a) The 2014 February 25 SGRE flux (>100 MeV) superposed on the Wind/WAVES dynamic spectrum showing the type II burst. Also shown is the >100 MeV proton flux as a proxy to the >300 MeV protons needed to produce the gamma-rays. (b) the associated CME in the STEREO-B spacecraft's SECCHI/COR2 and HI-1 fields of view. (c) the height-time plot of the CME leading edge measured along the white line in (b). The COR2 and HI-1 images correspond to the peak and end times of the SGRE event. When SGRE ended, the CME leading edge was at a heliocentric distance of ~66 Rs from the Sun's center. The yellow cross indicates a heliocentric distance of 70 Rs, which is likely to be the distance of the shock.

The CMEs underlying SGREs are all halo CMEs and their average speed exceeds 2000 km/s. The average ending frequency of type II radio bursts in SGRE events is ~200 kHz, implying that the shocks travel far into the IP medium. These characteristics of SGRE CMEs are shared by CMEs that produce ground level enhancement (GLE) in SEP events, indicating the high likelihood of the presence of >300 MeV protons needed to produce SGRE. Given that the average duration of SGRE events is ~11.3 hr (Gopalswamy et al. 2019a), a 2000 km/s CME would travel a distance of 116 Rs by the time the SGRE ends. The 2014 February 25 SGRE lasted for ~ 8.5 hours and the CME speed was ~2150 km/s (Gopalswamy et al. 2019b). It was possible to track the CME and shock using STEREO/COR2 and HI images. Figure 4 shows the CME in the STEREO-B/COR2 FOV at 01:54 UT and HI-1 FOV about 7.5 hrs later, when the SGRE ended. The CME leading edge was in the range 60-70 Rs at various position angles, confirming the large distance of the CME/shock over which >300 MeV protons were accelerated.

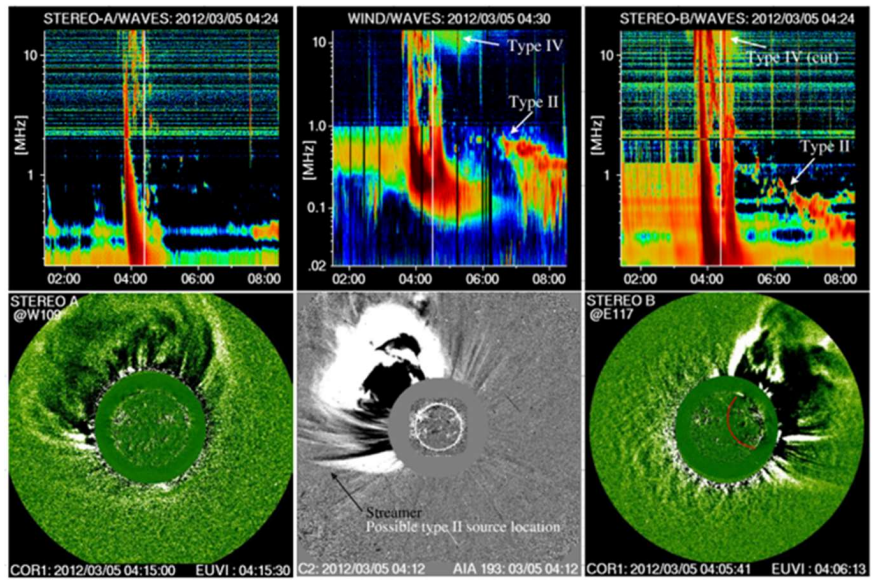

Figure 5. (upper panels) STEREO-A, Wind, and STEREO-B WAVES dynamics spectra. No type IV in STEREO-A (E161), full type IV in Wind (E52), and partial type IV in STEREO-B (W65) view. (Lower panels) CMEs in the three views are shown at the bottom (from Talebpour Sheshvan and Pohjolainen, 2018).

## 3. Low Frequency Type IV Bursts

Type IV bursts at frequencies below 15 MHz are rare extensions of metric type IV bursts and associated with high-speed (~1500 km/s on average) CMEs (Gopalswamy 2011; Hillaris et al. 2016). A new property of type IV bursts discovered recently is their directivity inferred from a combination of STEREO/WAVES and Wind/WAVES dynamic spectra (Gopalswamy et al.

2016): disk-center eruptions produce full (symmetric) type IV bursts, while limb-eruptions produce partial (asymmetric) type IV bursts. The symmetry is with respect to the time at which the type IV burst attains its lowest frequency. A related effect is the decrease in the number of type IV bursts with increasing central meridian distance of the associated eruption. The change in morphology of the type IV burst can be witnessed as an active region transits the solar disk. Based on multiview observations, Gopalswamy et al. (2016) concluded that the type IV emission is directed in a narrow cone with a half width <30º. The directivity has been attributed to the small angular extent of the post eruption arcades in which the type IV bursts originate compared to the larger angular extent of the associated CME. An alternative explanation has recently been proposed. In this scenario, the visibility of IP type IV radio bursts is affected by foreground shock-related high-density plasmas (Talebpour Sheshvan and Pohjolainen, 2018; Pohjolainen and Talebpour Sheshvan, 2020). These high-density plasmas could be located at the CME flanks, where the shock–streamer interactions can form type II radio bursts. At lower coronal heights these regions provide suitable conditions to observe the metric type IV bursts, as they are not located high enough to block the type IV emission over wide range of viewing angles. Figure 5 shows the type IV burst on 2012 March 5 from an eruption at N17E52. The type IV burst can be seen as a complete event in the Wind/WAVES spectrum, partially cut off in STEREO-B/WAVES spectrum, and no event at all in STEREO-A/WAVES spectrum. In STEREO-B and STEREO-A views, the source region is at N17W65 and N17E161 (backside) and hence the type IV emission region is viewed through plasma regions towards the flank. While the type IV emission intensity, i.e., directivity, depends on the source locations in different views, the underlying reason suggested by Talebpour Sheshvan and Pohjolainen (2018) is the blocking by compressed streamers at the flanks. Thus, the IP type IV bursts have important clues in understanding the overall eruption geometry including post eruption arcades (thought to be the type IV source) and the surrounding CMEs including streamers.

## 4. Diffuse Interplanetary Radio Emission (DIRE)

Gopalswamy et al. (2020b) reported on an interesting variant of type II bursts, called the diffuse interplanetary radio emission (DIRE) that originates from the flanks of a CME-driven shock. DIRE appears as a series of short duration bursts that resemble a type III storm but with a much narrower bandwidth (~3.5 MHz). Type III storms are associated with active regions outside of eruptions, while DIRE is associated with CMEs of moderate speed. The first reported DIRE event is from a polar CME associated with a filament eruption on 2000 April 18. Figure 6 shows another example of DIRE that occurred on 2002 March 10. The radio emission starts around 20:06 UT at the upper cutoff frequency of the Wind/WAVES instrument at ~14 MHz.

There were two CMEs around the time of the DIRE – a narrow one (width ~38º) from the southeast limb (average speed ~630 km/s; acceleration 1 m s$^{-2}$) and a wide CME (~90º) from the northwest (average speed ~720 km/s; acceleration -15.5 m s$^{-2}$). The first appearance time is the same in both cases (17:06 UT). We link the DIRE event to the northwest CME because of the favorable kinematics. The CME originated at a position angle (PA) a few degrees to the left of the brightest streamer in the northwest (PA~312º) (see Fig. 6a,b). By 17:30 UT, the CME extended from PA 294º to 354º (Fig. 6c). Around the time of the DIRE onset, the leading edge (LE) was at 13.6 Rs near the streamer, although the central PA is around 330º at a height of 12 Rs. Extrapolating the height-time plot to the DIRE onset (20:06 UT) gives a LE height of ~13 Rs. LASCO movies indicate that the right edge of the CME was partially overlapping with the streamer and the streamer was disturbed, but not disrupted. The speed was the highest in the region of streamer interaction. There was no activity reported in the online Solar Geophysical Data. A radio noise storm at 245 MHz was reported by the San Vito station of RSTN that started at 14:45 and ended at 16:19 UT, about four hours before the DIRE onset. The EUV disturbance in Fig.6c indicates that the CME must have originated from the limb or slightly behind and associated with a high-latitude filament eruption.

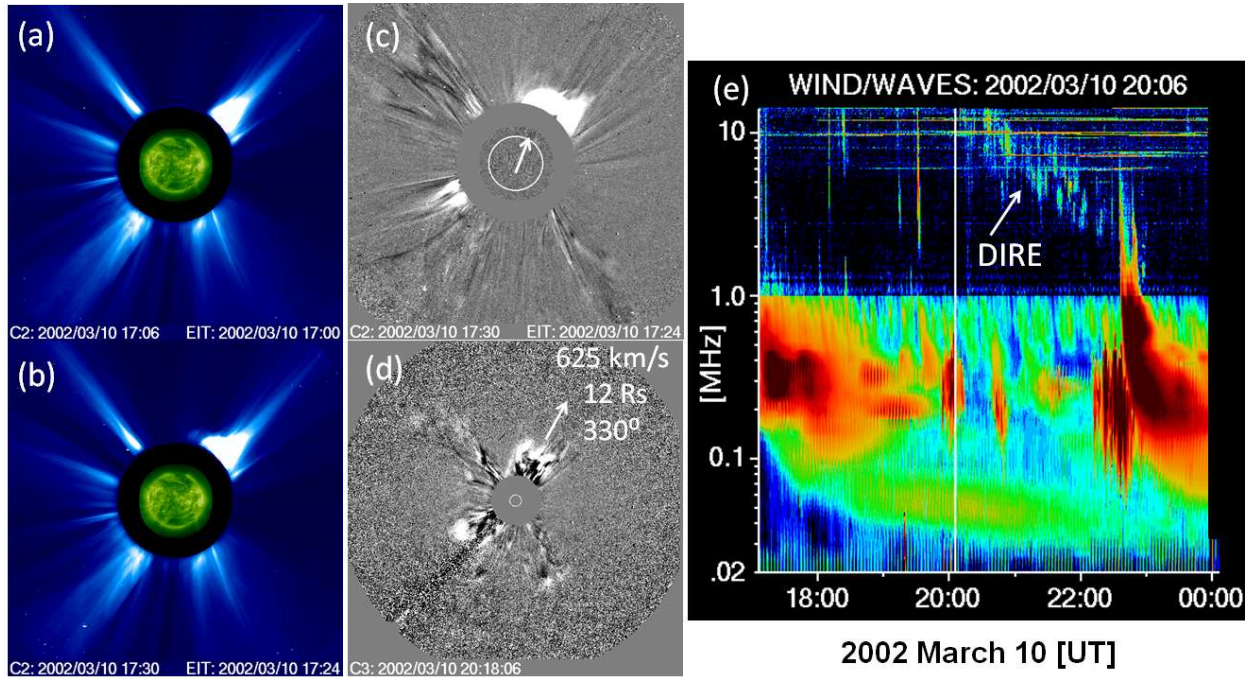

Figure 6. The 2002 March 10 CME observed by SOHO/LASCO (a-d) and the associated diffuse interplanetary radio emission (DIRE) observed by Wind/WAVES (e). The image at 17:06 UT (a) is subtracted from (b) to show the early stage of the CME (c) heading in the northwest direction with a weak EUV disturbance pointed by arrow. At 20:18 UT, just after the DIRE onset, the CME leading edge is already at ~13.7 Rs with a speed of 720 km/s. At the central position angle (CPA) of 330º, the CME height is ~12 Rs (d). The radio emission starts at the upper cutoff frequency of WAVES (~14 MHz) and the envelope drifts down to ~3 MHz at ~22:00 UT. The DIRE bandwidth is ~4 MHz, similar to the 2000 April 18 DIRE event.

The DIRE envelope drifts with a rate of ~0.08 MHz/s, which is higher than the typical drift rate of type II radio bursts around 14 MHz. The bandwidth is ~4 MHz, similar to the 2000 April 18 DIRE event. The emission frequency of ~14 MHz corresponds to a local plasma density of $2.4\times10^6$ cm$^{-3}$. Such a densities generally prevail at a heliocentric distance of ~2 Rs in the pre-event corona. The CME leading edge is at a height of 13 Rs at DIRE onset, which rules out the possibility of the emission coming from the nose region of the CME. The density in streamers is typically higher by a factor of ~5 than in the quiet corona, indicating that ~14 MHz plasma level can occur at a higher height in the streamers. The higher density in the streamer can lead to a lower Alfven speed, making it easier for shock formation and/or strengthening (Gopalswamy et al. 2004). Therefore, we suggest that the DIRE originates in the interaction region between the CME and the adjacent streamer. This was also shown to be the case during the 2000 April 18 DIRE. The DIRE envelope drifts with a rate df/dt ~0.08 MHz/s. For fundamental plasma emission at frequency f, the shock speed V is related to the drift rate by $V = 2L(1/f)(df/dt)$, where L is the density scale height. To get a shock speed of at least ~400 km/s near the DIRE onset (f=14 MHz), we need a scale height of L ~0.05 Rs. This can go up to 0.09 Rs, if the shock speed is ~720 km/s. In the typical upstream medium outside of streamers, the scale height is ~1 Rs or greater at frequencies at or below 14 MHz. The smaller scale heights are possible in the streamer stalks that the shock might pass through and produce the DIRE. The scale height obtained for the 2000 April 18 was also ~0.1 Rs. Thus, the DIRE events provide important clues to the understanding of CME-streamer interaction. It must be noted that the DIRE event is distinct from the flank emission we discussed earlier.

## 5. Type III Storms

Although eruption-related type III, type II, and type IV bursts are the most intense ones at DH and longer wavelengths, type III storms are long-lasting and provide information on the evolution of solar active regions. Type III storms are the low-frequency extensions of metric-wavelength type I storms and both contain large clusters of very short duration bursts (Fainberg and Stone 1970). The storms can last for many days as the source active region crosses the solar disk; the storm type III bursts can be observed at heliocentric distances of up to 170 Rs (Bougeret and Stone, 1984). Type III storms are known to show directivity (radial with a beam half width of ~40º, See Wright 1980). This results in the intensification of type III storms when the underlying active region crosses the central meridian. Like the intensity, the circular polarization of the type III storm bursts also peaks near the central meridian (Reiner et al. 2007). One of the interesting findings by Morioka et al. (2007) is that most of the active regions producing type III storms are located bordering active regions. Such a configuration leads to interchange reconnection of the active region field lines with open magnetic field lines in the neighboring coronal holes as the mechanism that produces nonthermal electrons responsible for type I and type III storms (Del Zanna et al. 2011). Morioka et al. (2015) proposed a variant of this scenario in which extremely large bipolar field lines reconnect with the active region field lines in order to explain the cutoff of the type III storm bursts at a frequency of ~100 kHz. Morioka et al. (2007) also reported that ordinary type III bursts from the same active region as type III storm bursts without affecting each other.

A large CME erupting from the source active region can disrupt the type III storm in progress, but the storm typically returns in several hours (Reiner et al. 2001; Gopalswamy 2016). Gopalswamy (2016) reported on the type III storm from AR NOAA 10720 that lasted for about a week. The storm started when the AR was at N12E10 (2005 January 14) ended at N14W61 (2005 January 20). The storm was repeatedly disrupted by five major CMEs from this region, but it recovered each time except for the last one: the storm disappeared after the January 20 eruption. Gopalswamy (2016) speculated that the STEREO/WAVES combined with Wind/WAVES can track the storms when they rotate behind the limb.

Here we briefly describe a storm from AR 12673 that was observed by both Wind/WAVES and STEREO/WAVES instruments and lasted for more than a month. This storm was also disrupted by at least three eruptions during its disk passage. The storm showed intensification near the central meridian and disruption by CMEs. The storm started around on 2017 August 27 around 22 UT as observed by STEREO-A/WAVES and ended on 2017 October 5 around 12 UT as observed by Wind/WAVES. The storm was observed by Wind/WAVES from August 31 around 17 UT to September 12 around 13 UT and then from October 1 at 23 UT to October 5 at 12 UT. The storm was not observed during three short intervals when the AR was near the limb in Wind and STEREO-A views for about three days when the source region was above the west limb (September 13-15) and for one day when it was near the east limb (August 30-31). In the second rotation, the storm was not observed when it went behind the west limb in STEREO-A view on September 27, until it was picked up by Wind/WAVES on October 1.

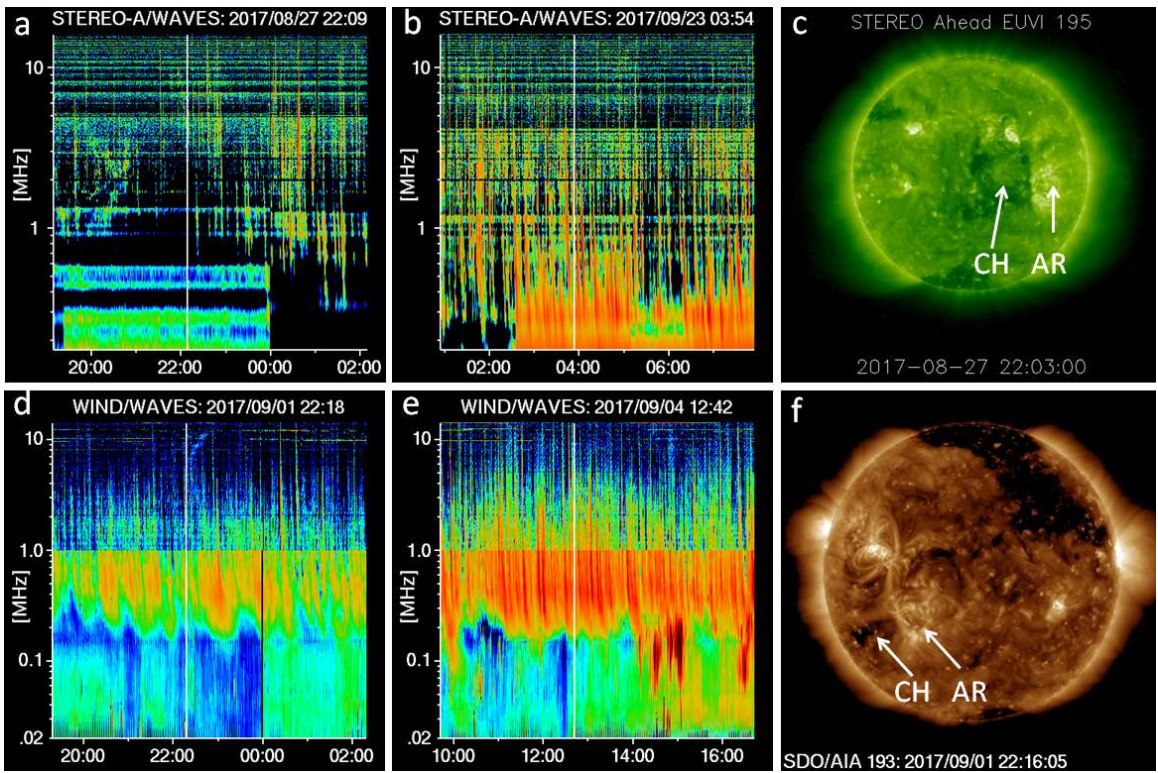

Figure 7. upper: Type III storm in the radio dynamic spectra near its onset (a) and maximum (b) from STEREO A along with the location of the source active region (AR) and nearby coronal hole (CH) in the EUV image taken near the storm onset (c). lower: the corresponding dynamic spectra from Wind/WAVES (d,e) and the source region from SDO/AIA (f). The red and blue colors correspond to the highest and lowest burst intensities above background.

The dynamic spectra in Fig. 7 show the storm at two instances, one near the onset and the other near the peak. The location of the underlying active region and the adjacent coronal hole are also shown in EUV images taken by STEREO-A at 15 Å and SDO/AIA at 193 Å. The storm onset can be seen in Fig. 7a first starting at higher frequencies and then appearing at all frequencies. At the onset on August 27, the active region was located at ~S10W36 in STEREO-A view. Even though the active region was present for several days earlier, there was no storm. The storm was not visible beyond August 30 when the region reached the west limb in STEREO-A view. In the Wind/WAVES dynamic spectrum the storm started at the end of August 31 when the region was at E36 in Earth view (Fig. 7d). The storm was very intense when the region was near the central meridian on September 4 (Fig. 7e). The storm was no longer observed by Wind/WAVES when the active region rotated behind the west limb in Earth view after September 10. The storm was again observed by STEREO-A on September 15 when the region rotated into STEREO-A FOV (around E65 in STEREO-A view). The storm intensified when it crossed the central meridian in STEREO-A view (Fig. 7b) before weakening on September 26 and ending on September 27 when it rotated behind the west limb in STEREO-A view. When the active region reached W19 in Earth view on October 1, the storm was observed by Wind/WAVES for the second time before ending on October 5 (region at W75). These observations illustrate that there is an east-west asymmetry: the storm is observed better from the western hemisphere (even from behind the limb), but the region needs to be away from the east limb. It is possible that the Parker spiral along which type III electrons propagate and the directivity of the radio emission may account for this asymmetry.

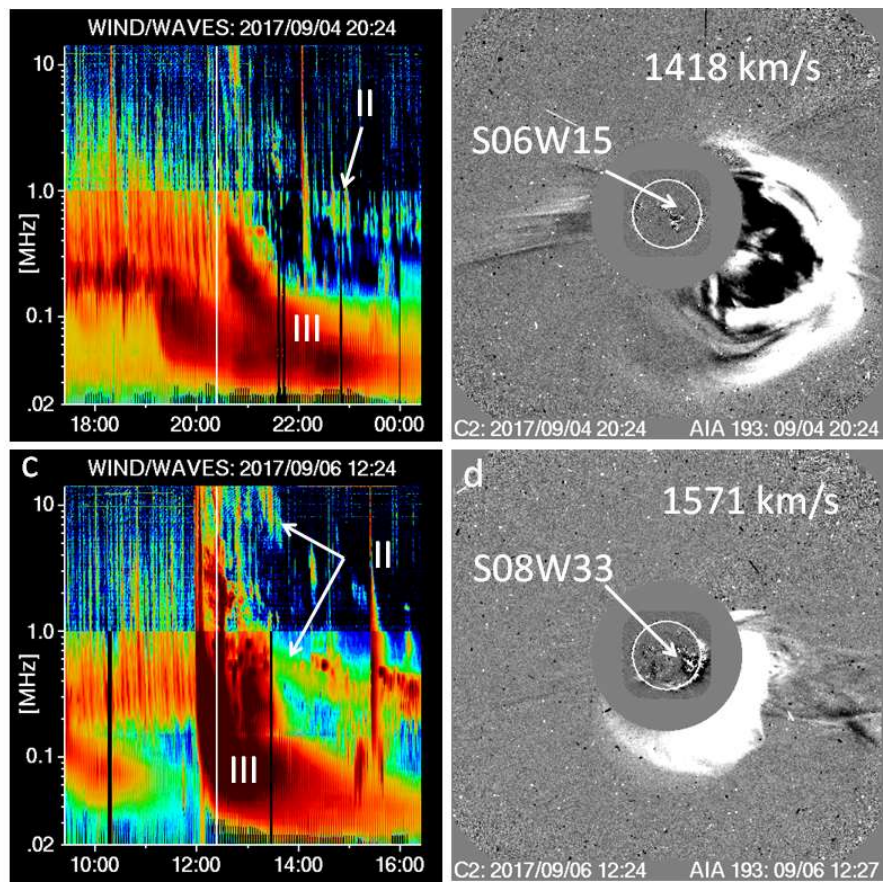

Figure 8. Solar eruptions that disrupted the type III storm on September 4 and 6. The eruption type II and Type III bursts are marked as II and III, respectively. The type III storm consists of fine vertical structures in rapid succession to the left of the eruption type III bursts. The vertical white lines on the Wind/WAVES dynamic spectra (a,c) indicate the times of the CMEs on the right (b,d). The locations of the eruptions and the CME speeds are marked in b and d. The temporary disappearance of the type III storm after the eruption type III bursts is obvious.

There were three major solar eruptions from AR 12673 on September 4, 6, and 10. These events have received a lot of attention because they are all associated with large SEP events including a GLE event, two SGRE events, and a large geomagnetic storm (Gopalswamy et al. 2018a,b; Bruno et al. 2019; Struminsky 2019). Wind/WAVES had a data gap when the September 10 eruption happened, but the storm was intense just before the eruption and was considerably weak when Wind/WAVES observations resumed on September 11. The other two eruptions had a clear impact on the storm as illustrated in Fig. 8. The September 4 eruption had a slower CME, but still fast enough to produce a large SEP event. The September 6 event had a faster CME and produced a large SEP event that was associated with an SGRE event. The eruption-related type III and type II bursts are very intense compared to the storm type III bursts. The eruption type III burst marks the time the type III storm is temporarily disrupted. The storm recovered after ~4 hrs following the September 4 disruption, and it took 8 hrs after the September 6 disruption. The recovery time is important parameter that gives clues to understand the how the active region – coronal hole system relaxes back after a flux rope leaves the active region. These observations are not in agreement with the report that the storm bursts and ordinary type III bursts coexist without affecting each other. It is quite possible that the ordinary type III bursts originated in a neutral line in the same active region but different from the edge where the storm bursts are produced.

Throughout the period the storm was in progress, the coronal hole was positioned adjacent to the active region. Thus, the active region – coronal hole configuration (Fig. 7c,f) is similar to what Morioka et al. (2007) found and consistent with the interchange reconnection scenario proposed by Del Zanna et al. (2011). The combination of STEREO and SOHO coronagraph data with the radio data from STEREO and Wind has provided a unique opportunity to understand the type III storms and establish their properties much better.

## 6. Summary

We focused on some recent results in low-frequency solar radio physics that illustrate the enormous progress made in understanding nonthermal radio emission from the inner heliosphere that include both eruptive and non-eruptive energy releases. The direction-finding technique from STEREO/WAVES has started contributing to the understanding the origin of type II bursts on the shock front: nose or flanks or both. This technique needs to be used on different burst types for a better understanding of their origin in solar eruptions. Low-frequency Type IV bursts, the rarest among the three eruption bursts have introduced new opportunities to investigate the source and surrounding magnetic structures including CMEs, shocks, and streamers because of the observed directivity. The newly discovered DIRE events seem to be a variant of type II bursts but differ in morphology and drift rate. More investigation is needed to fully understand these bursts especially because they occur alone or with regular type II bursts, located at shock flanks interacting with streamers. Solar gamma-ray events lasting beyond the impulsive phase have been found to be quantitatively related to IP type II bursts. This provides an opportunity to find why not all IP type II bursts are associated with such gamma-ray events to determine additional factors that affect gamma-ray emission. The combination of STEREO/WAVES and Wind/WAVES provide better longitudinal coverage of type III storms. Since these storms are due to energetic electrons energized in solar active regions outside eruptions, they can provide important information on the high-altitude coronal structures overlying solar active regions and their interaction with neighboring open field structures. The disruption of type III storms by solar eruptions is a useful indicator of the source region of the storm. The disruptions and the subsequent recovery of the storms provide important clues to the timescales over which the corona recovers from an eruption.

**Acknowledgments.**

I appreciate the opportunity to speak during the 80$^{th}$ birthday celebrations of Dr. H. S. Sawant. I thank the SOHO, STEREO, SDO, and Wind teams for making their data that helped in obtaining the results presented in this paper. Work supported by NASA's LWS TR&T program.